\documentclass[namedreferences]{solarphysics}
\usepackage[optionalrh]{spr-sola-addons} 
\usepackage{graphicx}                    
\usepackage{color}                       
\usepackage{url}                         

\begin{document}

\begin{article}

\begin{opening}

\title{Detection of acceleration processes during the initial phase of the 12 June 2010 flare}

%
\author{L.K.~\surname{Kashapova}\sep
 N.S.~\surname{Meshalkina}\sep  M.S.~\surname{Kisil} }

%

%
  \institute{$^{1}$ Institute of Solar-Terrestrial Physics RAS SB, P.O. Box 291, 664033 Irkutsk, Russia
                     email: \url{lkk@iszf.irk.ru-a} email: \url{nata@iszf.irk.ru}
             }

\begin{abstract}
We present an analysis of the plasma parameters during the initial phase of the 12 June 2010 flare (SOL2010-06-12T00:57). A peculiarity of the flare was the detection of $\gamma$--emission that is unusual for such weak and short event. The analysis revealed the presence of a flare precursor detected about 5 minutes before the flare onset in 94~\AA \ images which spatially coincided with the non-polarized microwave (MW) source at 17 GHz (\textit{the Nobeyama Radio Heliograph}) that is the Neutral Line associated Source (NLS). A comparison of the results obtained from MW data by \textit{the Nobeyama Radio Polarimeters} and \textit{the multi-frequency Siberian radioheliograph} (the new 10-antenna radio heliograph prototype at 4.6 and 6.4 GHz) and hard X-ray (HXR) observations by \textit{the Fermi Gamma-ray Space Telescope} reveal the presence of accelerated electrons during the flare initial phase. The analysis of MW and HXR spectra also confirms the presence of accelerated particles. Moreover a good temporal correlation between several lightcurves in different HXR energy bands and at MW frequencies indicates generation of both HXR and MW emission by a common population of accelerated electrons.
Detection of accelerated particles during the initial phase of the flare and soft-hard-harder (SHH) behavior of the spectra indicate several episodes of particle acceleration and confirm the non-impulsive type of the flare evolution.
\end{abstract}

%

\end{opening}

\section{Introduction}
     \label{S-Introduction}
     The presence of a flare precursor has attracted the attention of researchers for tens of years. Nowadays, we distinguish between a precursor and the early phase of the flare. Precursors precede the impulsive phase of a flare and they can be seen from UV to soft X-rays (SXR) as a small-scale brightening happening up to some tens of minutes before the flare impulsive phase. \cite{2011SSRv..159...19F}. Flare precursors often do occur nearby but not usually at exactly the same location as  the forthcoming flare or where most of the flare emission will subsequently occur (for example, \opencite{1996SoPh..165..169F}; \opencite{2003A&A...399.1159F}; \opencite{1998SoPh..183..339F}; \opencite{2001ApJ...560L..87W}).

An increased thermal emission preceding the onset of the
impulsive phase is one of the important characteristics of the initial (or early) phase of flare evolution. Of course, this is irrelevant to unusual "cold" flares, which exhibit significant non-thermal emission without any noticeable thermal emission \cite{2011ApJ...731L..19F}. The analysis of a large number of flares observed simultaneously in HXR, SXR, and H$\alpha$ by \inlinecite{2002SoPh..208..297V} showed that more than 90\% of the studied cases presented an increase in SXR emission before the impulsive phase.
\inlinecite{1987SoPh..107..271L} based on models employing current sheets (e.g., \opencite{1977ApJ...216..123H}) assumed that the pre-flare phase was triggered by new emerging magnetic flux. The proposed types of reconnection regimes could be used to explain of both the pre-flare and impulsive phases.
\inlinecite{2006ApJ...637..522W} demonstrated that the multi-thread hydrodynamic
model could also reproduce the previous SXR emission.

 Another mechanism  explaining  the pre-flare emission is con\-duc\-tion driven evaporation. The model that could explain the emission from the low-tem\-pe\-ra\-ture regions of flares was firstly developed theoretically by, e.g. \inlinecite{1974SoPh...39..415S}. The presence of the conduction that drives chromospheric evaporation during different phases of flare was confirmed by observations \cite{1988ApJ...329..456Z,1982AdSpR...2..145C}. \cite{1988ApJ...329..456Z,1982AdSpR...2..145C}. \inlinecite{2009A&A...498..891B} compared the contribution of beam-driven evaporation and conduction-driven evaporation to the HXR emission during the pre-flare phase of the flares. They demonstrated that  the observations of the  quasi-thermal HXR emission was compatible with chromospheric evaporation driven by a saturated heat flux.
This result is supported by
the recent spatially resolved observations of chromospheric heating carried out with \textit{the Solar Dynamics Observatory} (SDO) and \textit{The Reuven Ramaty High-Energy Solar Spectroscopic Imager} (RHESSI)~\cite{2011A&A...533L...2B,2011ApJ...734L..15K,2012A&A...540AB}.

Another explanation to the presence of thermal emission before non-thermal one is that the emission could be below the detection threshold of HXR at the beginning of the flare \cite{1988SoPh..118...49D}.
\inlinecite{2009ApJ...705L.143S} showed that
the energy deposited by non-thermal electron beams is sufficient to heat flare loops to temperatures in which they emit SXR closely following the GOES light curve. The authors noted that RHESSI spectra indicated the presence of non-thermal
electrons also before the impulsive phase. They assumed that the dominant energy transport mechanism during the rise phase of solar flares is electron-beam driven evaporation and used the parameters derived from RHESSI spectra for the non-thermal electron beam, which they consider as the heating source in a hydrodynamic model of the analyzed flare. These results are in agreement with those of the study of the pre-flare event by \inlinecite{2011MNRAS.411.1562Z}. Unlike the event considered by \inlinecite{2009ApJ...705L.143S}, this flare had high enough HXR flux in the 25--50 keV energy band. Results of the modeling also revealed that the calculated temperature closely agreed with the value of the electron temperature derived from HXR spectrum. So, the most discussed problems concerning the initial or early stages of flare evolution are mainly related to the mechanisms working during the thermal pre-flare phase and the role of accelerated particles in this process.

An M2.0 white-light flare occurred in active region 11081 (N22W45) on 12 June
2010. The HXR light curve of this event was very impulsive. \inlinecite{2011SoPh..269..269M} noted that its duration (at half
maximum in 50 keV) was about 25 seconds. However it was the first $\gamma$--ray flare of the current Cycle 24. It is unusual that such a weak event had a spectrum reaching such high energies \cite{2009ApJ...698L.152S}.
\inlinecite{2011SoPh..269..269M} studied this event using white-light high-resolution imaging spectroscopy, including continuum, Doppler, and magnetic signatures in the photospheric
Fe I line at 6173.34~\AA \ and its neighboring continuum. During the impulsive phase,
a bright white-light kernel appeared in each of the two magnetic footpoints. The Fe I
line appeared to be shifted to the blue. The authors did not detect a seismic wave from
this flare as predicted by \inlinecite{1972ApJ...176..833W} and observed by \inlinecite{1998Natur} and others, which is also unexpected for such event. \inlinecite{2011SoPh..269..269M} attributed  the peak of HXR flux preceding the main
impulsive burst to the pre-flare phase. No information was obtained from \textit{the Helioseismic and Magnetic Imager}(HMI) continuum and Doppler images. RHESSI did not detect any significant flux above 100~keV.

Results of previous investigations of the  precursor and the early phase of the flare indicate the possible  keynote role of these phenomena in the flare scenario.
We studied the initial phase of the event considering the classical indicators of accelerated electrons -- MW and HXR emission. The main objective was to confirm the presence or absence of electron acceleration and  to find an explanation for several unusual characteristics.

\section{Observations}
     \label{S-Observations}
The 12 June 2010 flare was the first flare observed by the prototype of \textit{the multi-frequency Siberian radioheliograph} \cite{Prototype} of the Radio Observatory of the Institute of Solar Terrestrial Physics located in Badary (Russia). At that time the prototype was a 10-antenna interferometer and observed at two frequencies. The MW flux was obtained at two frequencies (4.6 and 6.4 GHz) with a temporal resolution of 0.56 sec. \textit{The Nobeyama Radio Polarimeters }(NoRP) data at 1, 2, 3.75, 9.4, 17, 35, and 80 GHz were used in our study. The relative calibration of the fluxes of the 10-antenna prototype was carried out using the values of the quiet Sun obtained by NoRP \cite{torii,shibasaki,nakajima}.
The standard time resolution of  NoRP data is 1~sec, and for flares 0.1~sec.
\textit{The Nobeyama Radioheliograph} (NoRH, \opencite{nakajima94}) produces images at 17 and 34 GHz.
We used images of intensity and polarization  at 17 GHz taken from NoRH. The HXR flux evolution and spectra were obtained from \textit{the Gamma-Ray Burst Monitor} (GBM) of \textit{the Fermi Gamma-ray Space Telescope} \cite{Meegan2009ApJ}. The line-of-sight magnetograms and continuum images for the white-light flare kernel subtraction were obtained by HMI/SDO \cite{HMI}.  We also used EUV images by t\textit{he Atmospheric Imaging Assembly} (AIA)\textit{}/SDO \cite{2011SoPh..tmp..316B} in our analysis.

\subsection{Evolution of MW and HXR fluxes }
     \label{Fl-Observations}

\begin{figure}
\begin{center}
\parbox{0.47\hsize}{
\resizebox{ \hsize}{!}{ \includegraphics{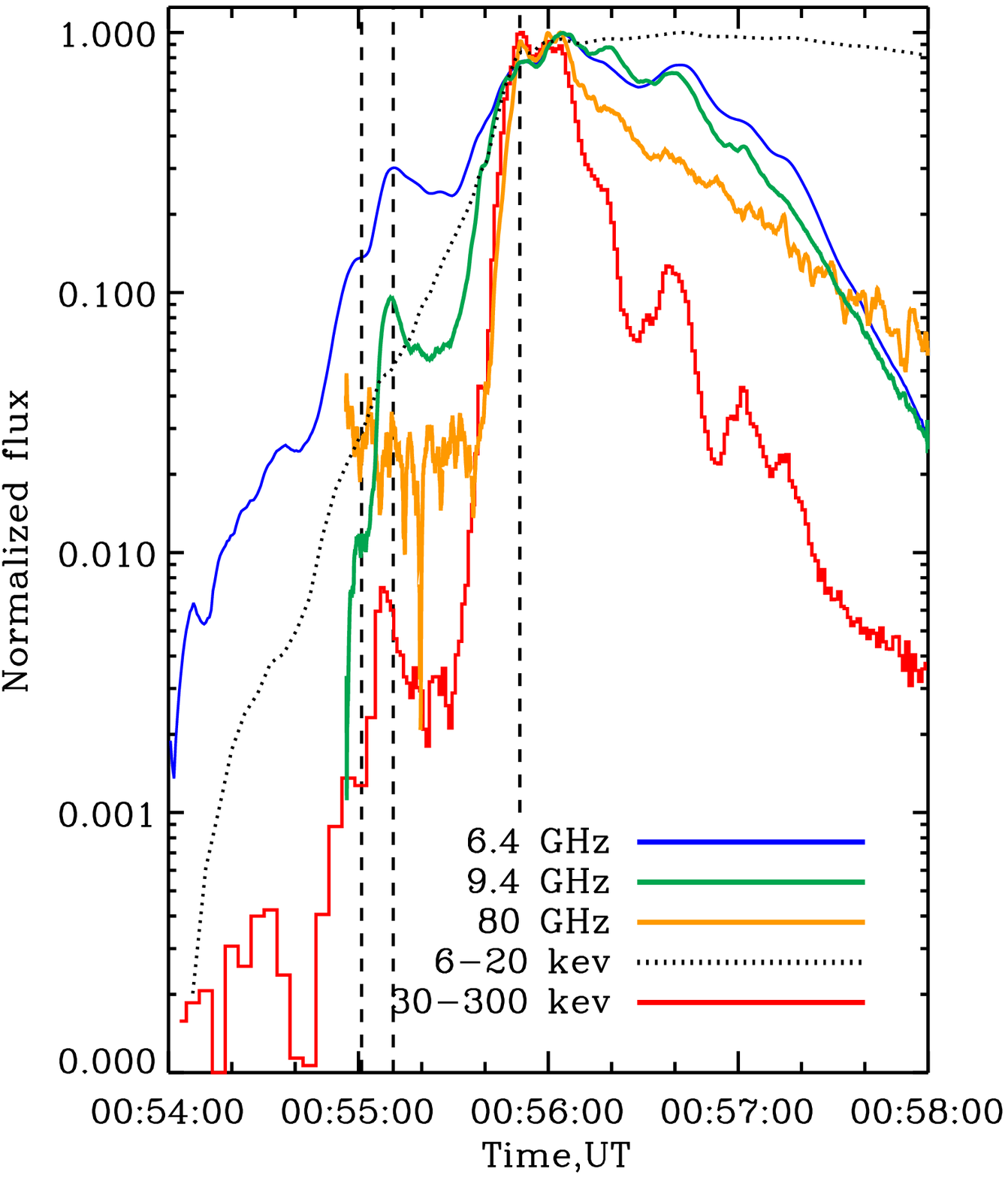}}}
\parbox{0.47\hsize}{
\resizebox{ \hsize}{!}{ \includegraphics{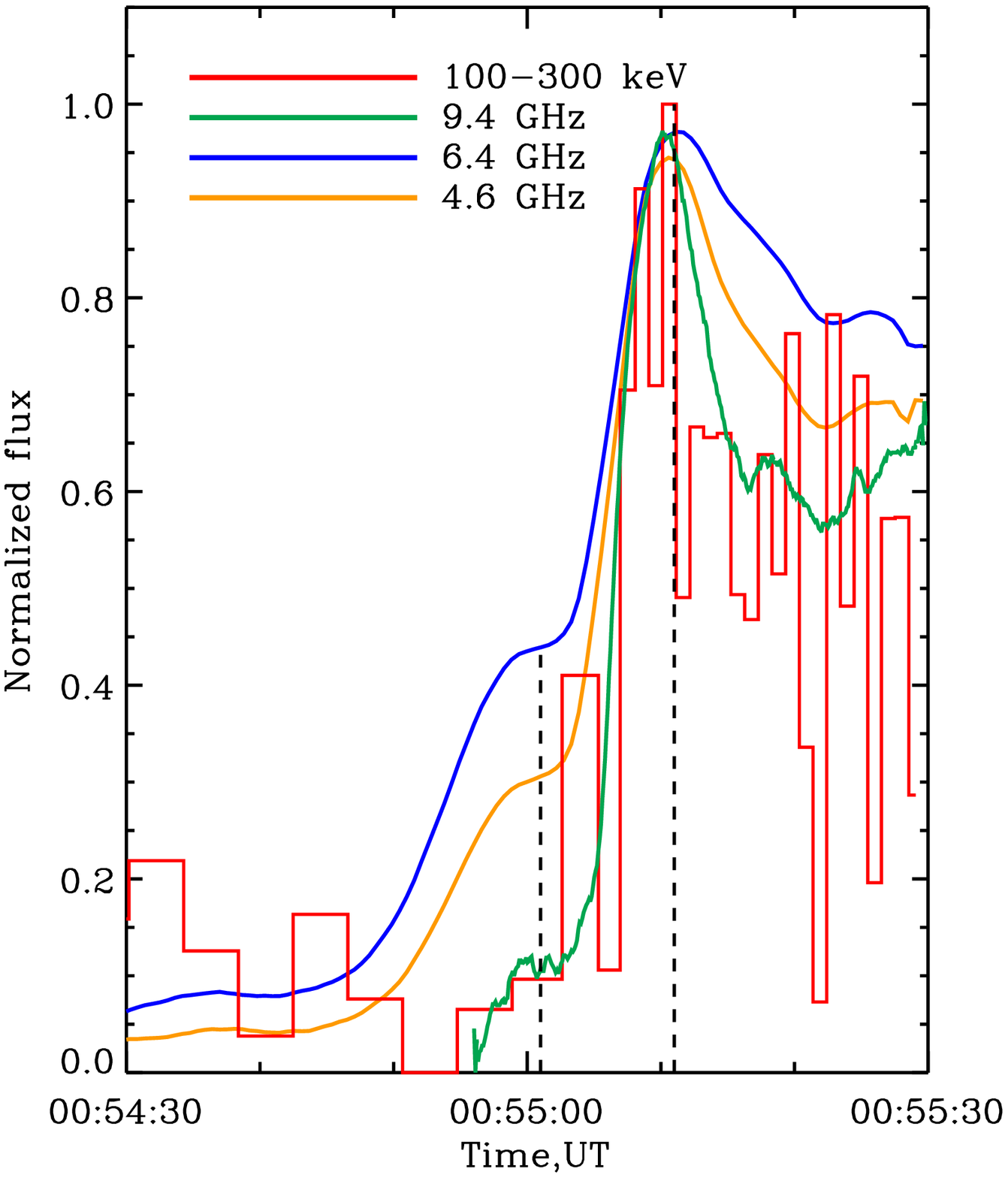}}}

\caption{Evolution of MW flux and HXR fluxes during the flare. The MW flux at 4.6 and 6.4 GHz is obtained by the 10-antenna radio heliograph prototype, the MW flux at 9.4 and 80 GHz is obtained by NORP. The left panel shows the evolution during the whole event. Vertical dashed lines correspond to 00:55:01~UT, 00:55:10~UT and 00:55:51~UT (the main peak). The right panel shows the evolution of the MW flux at 4.6, 6.4 and 9.4 GHz and the HXR flux at 100-300~keV  during the early phase of the flare. }

  \label{fig:0}
\end{center}
\end{figure}

 \inlinecite{2011SoPh..269..269M} defined  the pre-flare phase initiation  at 00:54:11 UT and the brightening onset at 00:54:41~UT. Analyzing the data from the \textit{Extreme Ultraviolet Variability Experiment} (EVE) in 304~\AA, \inlinecite{2011SoPh..tmp..362H} defined the pre-flare phase from 00:50:09 to 00:54:29~UT and the impulsive phase of the flare from 00:55:59  to 00:56:49~UT. So, we will consider the period  00:54:30--00:56:00~UT as the initial or early phase of the flare. The activity before this period is the pre-flare phase or precursor.

 As one can see in Figure~\ref{fig:0} (left panel), the MW time profiles at 6.4 and 9.4 GHz are in good agreement  with the evolution of the HXR flux from the onset to the maximum of the flare. During the first 40--50  seconds a better correlation is seen between 6.4 GHz and the HXR flux in 6--20 keV. After that, we clearly see a peak in HXR in the energy band 30--300 keV that corresponds  to the non-thermal phase. The close temporal correlation between the 30--300 HXR and MW fluxes is seen from about 00:55:50~UT. \inlinecite{2011SoPh..tmp..362H} did not find any significant HXR flux above 100 keV during the initial phase of this flare while the 304~\AA\ emission started to rise. As one can see in in their Figure 5, the EUV flux growth began at the time corresponding to both MW and HXR flux increase; the first peak in 30-300 keV also occurred around that time.

We compared the evolution of the MW flux in 4.6, 6.4 and 9.4 GHz to the 100--300~keV HXR flux (Figure~\ref{fig:0}, right panel). The fluxes showed the same good agreement as it was found for the wider HXR energy band (30--300 keV). So, these results support the conclusions by \inlinecite{1988SoPh..118...49D} and \inlinecite{2009ApJ...705L.143S} that accelerated electrons are present in the early phase of the flare but their flux could be at or below the detection threshold.  In any case, we note that  MW and the non-thermal HXR fluxes showed a close correlation that indicates the presence of accelerated electrons during the early phase of this event and that both HXR and MW emissions were produced by a common electron population.
Moreover, Figure~\ref{fig:0} indicates that microwaves lag behind hard X-rays, most likely, due to trapping of microwave-emitting electrons (trapping effect), while hard X-rays are emitted by electrons directly precipitating in the dense of the solar atmosphere.

\subsection{Spectral parameters }
     \label{SP-Observations}
\begin{figure}
\begin{center}
\parbox{0.48\hsize}{
\resizebox{ \hsize}{!}{ \includegraphics{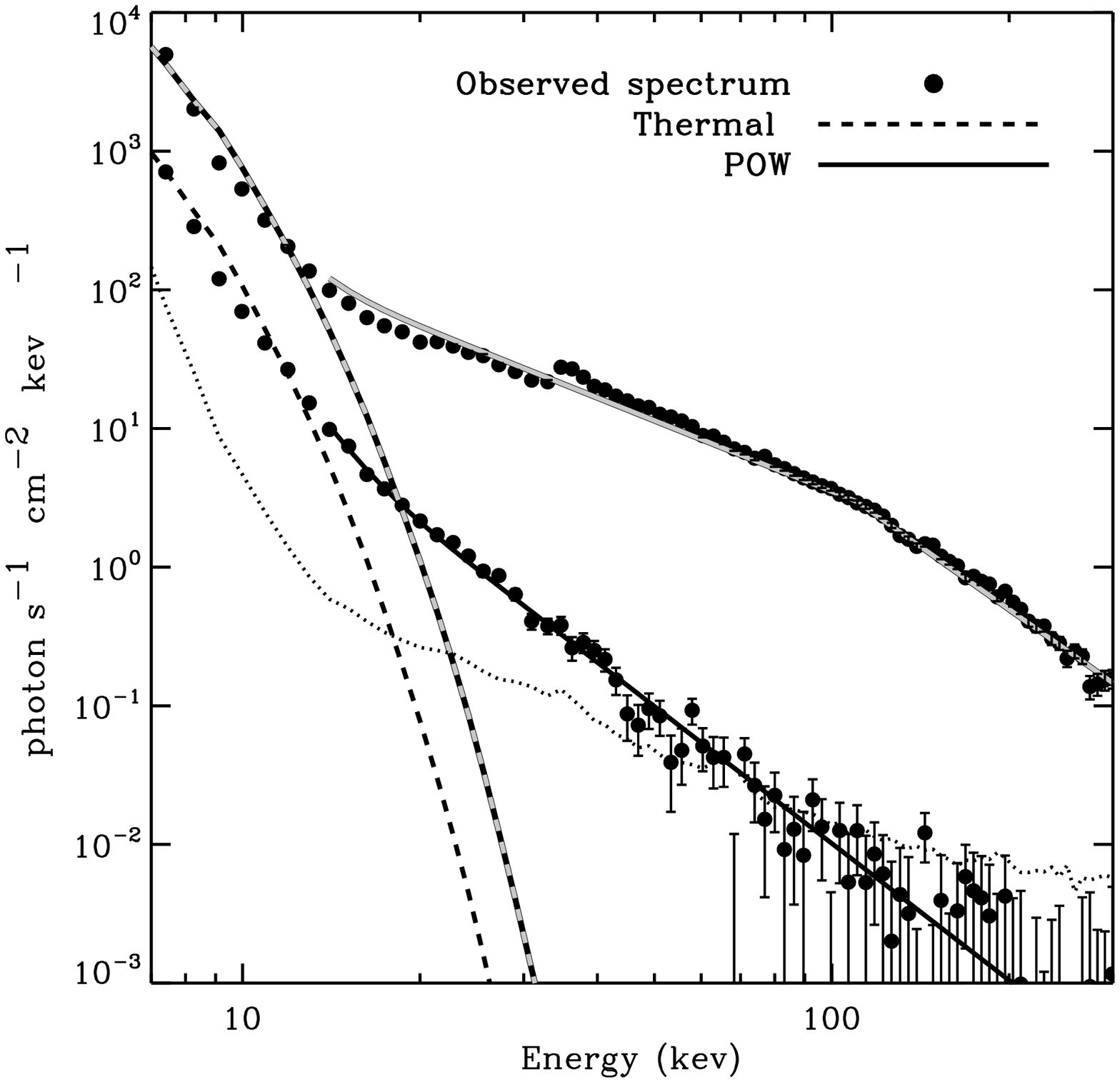}}}
\parbox{0.48\hsize}{
\resizebox{ \hsize}{!}{ \includegraphics{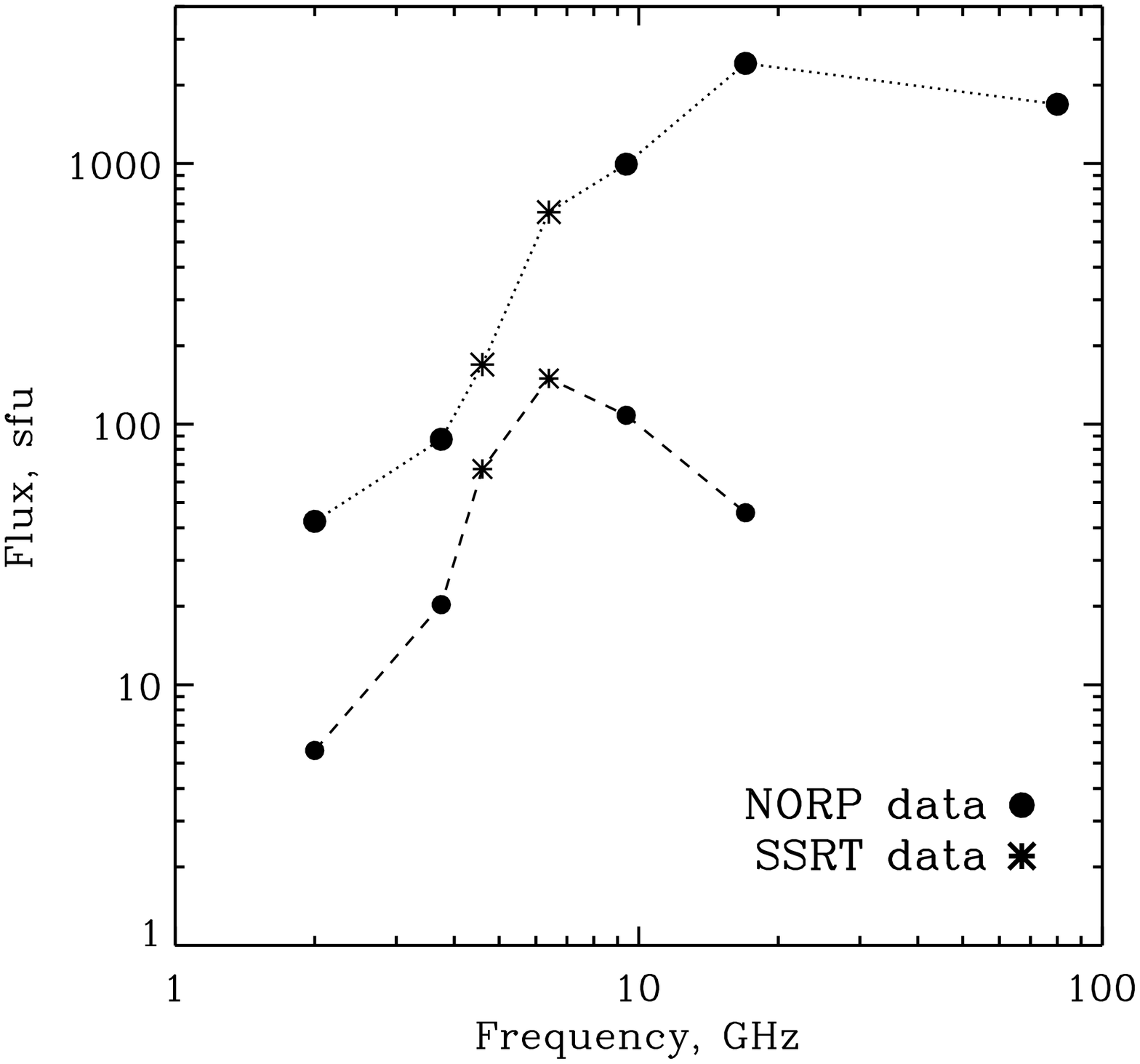}}}

\caption{ Left panel: the HXR spectra by FERMI at 00:55:10.7~UT (here and after, the time refers to the mean time of the interval along  which the spectrum is accumulated; this is marked by the black line) and 00:55:50.6~UT (the grey line). The dotted line shows the background flux at 00:55:50.6~UT. The dashed lines show the fitted thermal component and the solid lines show the fitted broken power-law component(POW). Right panel: observed MW spectra combining NORP data and the observations of the 10-antenna radio heliograph prototype (SSRT) at 00:55:10~UT (the dashed line) and 00:55:51~UT (the dotted line).}

  \label{fig:1}
\end{center}
\end{figure}

In studying different phases of the event, we fit the HXR spectra observed by FERMI with a model consisting of a thermal and broken power-law components.
The albedo correction was done using the approach
by \inlinecite{2006A&A...446.1157K} incorporated into the the SPectrum EXecutive (SPEX) software \cite{2002SoPh..210..165S}. This approach that takes into account Compton backscattering and photoelectric absorption using Green functions was derived by \inlinecite{1995MNRAS.273..837M} and applied for the first time in solar physics by \inlinecite{2006A&A...446.1157K}. This approach allows to
correct any X-ray spectrum and can be used for both forward and
inverse spectrum calculations. The coefficient  accounting for the anisotropy of the X-ray source was taken in the Eddington  approximation as the ratio of the flux toward the Sun and the flux toward the observer \cite{2006A&A...446.1157K}. We assumed that the emission source was isotropic and the ratio was equal to 1.

One important parameter of the broken power-law  approach is the broken point energy. In our fitting it was a free parameter with an initial value of 100 keV. The fitting of the initial phase of the flare revealed that  the contribution of the non-thermal portion of the HXR spectrum was significant. At this time the non-thermal part is better described by a single power-law. The spectrum for the time interval 00:54:58.9--00:55:03.0~UT indicated an electron temperature (TE) of 15 MK and an emission measure (EM) of about 10$^{48}$cm$^{-3}$. The photon spectral index $\gamma$ was 5.7. The spectrum for the time interval 00:55:10.1--00:55:11.2~UT (the first peak in HXR in the energy range 30-300~keV) showed TE=19~MK and  EM=8$\times$10$^{47}$cm$^{-3}$. The photon index $\gamma$ = 3.3 showed a fast hardening.

The spectrum determined for the interval 00:55:50.1--00:55:51.1~UT (the beginning of impulsive phase) showed a prominent feature at 33 keV. As one can see in Figure~\ref{fig:1}, the spectrum of the background flux also showed this prominent feature at the same energies. \inlinecite{Meegan2009ApJ} had noted existence of such feature arising from the K-edge in iodine in FERMI spectra. To be sure about the instrumental origin of this feature, we looked at the spectrum obtained by RHESSI \cite{Lin02} for the same time. The RHESSI spectrometer uses germanium detectors while the GBM/FERMI has sodium iodide (NaI) detectors \cite{Meegan2009ApJ}. The necessary correction of the pile-up and the albedo effects was applied to the spectrum obtained by RHESSI. We did not find any jumps near 30 keV for these spectral data. So we concluded that the observed feature is the instrumental feature mentioned by \inlinecite{Meegan2009ApJ}.

The non-thermal part of the spectrum at 00:55:50~UT showed a double power-law shape. The fitting gave $\gamma_{1}$=1.7 and $\gamma_{2}$=3.0 (the photon spectral indices above and below the break point in energy, respectively) and the energy break was at about 110~keV. This double-power law shape could be explained in different ways (including the effect of the HXR emission anisotropy on the albedo correction \cite{2007A&A...466..705K}) but, in any case, we observe the hardening of the spectral index. The thermal parameters of the flare plasma for this phase are TE= 22~MK and EM=3$\times$10$^{48}$cm$^{-3}$. They also indicated some increase in the temperature and the emission measure but these changes are not so tremendous as the changes in the parameters characterizing an acceleration process.

The MW spectra (Figure~\ref{fig:1}, right panel) could be explained by the gyrosynchrotron mechanism. In the initial phase (00:55:10~UT) the peak frequency was at 6.4 GHz and during the impulsive phase, it shifted to higher frequencies. Because of the absence of reliable measurements at 34 GHz, we can say that the peak frequency was equal to or higher than 17 GHz.

In order to compare the spectral indices of the microwave-emitting electrons and the HXR-emitting ones, we convert the photon spectral indices inferred from HXR and MW spectra to electron spectral indices. For microwave emission, the electron spectral index $\delta$ is calculated from the photon spectral index $\alpha$ as $\delta=1.11\alpha+1.36$.

Studying the correlation between MW and HXR spectral indices, \inlinecite{2000ApJ...545.1116S}
found that non-impulsive flares display a more significant difference between HXR and MW spectral indices  than impulsive ones. These authors statistically estimated the differences between the spectral indices for impulsive and non impulsive events. \inlinecite{2000ApJ...545.1116S} converted the photon spectral indices $\gamma$ obtained from HXR spectra to the electron spectral indices $\delta_{HXR}$  as $\delta_{HXR}$=$\gamma+1.5$. In order to apply the results of these authors to our study, we had to use the same formula to calculate $\delta_{HXR}$.

 The electron spectral indices were $\delta_{MW}$=2.9 and $\delta_{HXR}$=4.8  at 00:55:10~UT. We found $\delta_{MW}$= 1.6 and $\delta1_{HXR}$=3 or $\delta2_{HXR}$=4.5 at 00:55:50~UT. We note that the application of the thick-target model gave values of the electron spectral indices that were close to the values obtained based on the assumption $\delta_{HXR}$=$\gamma+1.5$.
Thus the electron spectral index obtained under the assumption of a thick-target model was  4.8 at 00:55:10~UT and the electron spectral indices at  00:55:50~UT were  2.3 and 4.4, correspondingly.
 The fact that the electron spectral index taken from MW data is harder than the electron spectral index obtained from HXR observations was noted by \inlinecite{1994ApJS...90..599K}. There are several models that explain such difference (see for example, \opencite{2000ApJ...545.1116S}). The difference between both spectral indices was about 2 at 00:55:10~UT and exceeded 1 at 00:55:51~UT. Following the results of \inlinecite{2000ApJ...545.1116S} a flare with such characteristics is non-impulsive. According to these authors, non-impulsive flares are  events that present a gradual rise and decay of the flux with a burst duration of more than 2 minutes. They proposed that non-impulsive flares are most likely the result of a prolonged injection of accelerated electrons or trapping effect. In this case, the electron trapping would result in the accumulation of higher energy electrons in the MW source and a harder MW electron spectral index  with respect to the HXR index (e.g. \opencite{1994R&QE...37..557M}).

The peak duration of the studied flare indicates its impulsive character, but the difference between the electron indices taken from MW and HXR points to a non-impulsive character. As it was noted above a non-impulsive flare should be followed by trapping which is clearly seen in Figure~\ref{fig:0}. So, in spite of its short duration, we could characterize the energy release process as non-impulsive. In this case the difference between MW and HXR indices could be explained by the so called "trap plus precipitation" model. The trapped electrons emitting in microwaves lose the lower energy electrons due to  the energy-dependent Coulomb collisions (e.g. \opencite{1984AdSpR...4..153T}). So, MW emission is generated by these electrons is harder than the HXR emission that is produced by the precipitated electrons keep the original injection spectrum, which is softer.

\subsection{Evolution of the emission sources.}
     \label{Im-Observations}

\begin{figure}
\begin{center}
\parbox{0.32\hsize}{
\resizebox{ \hsize}{!}{ \includegraphics{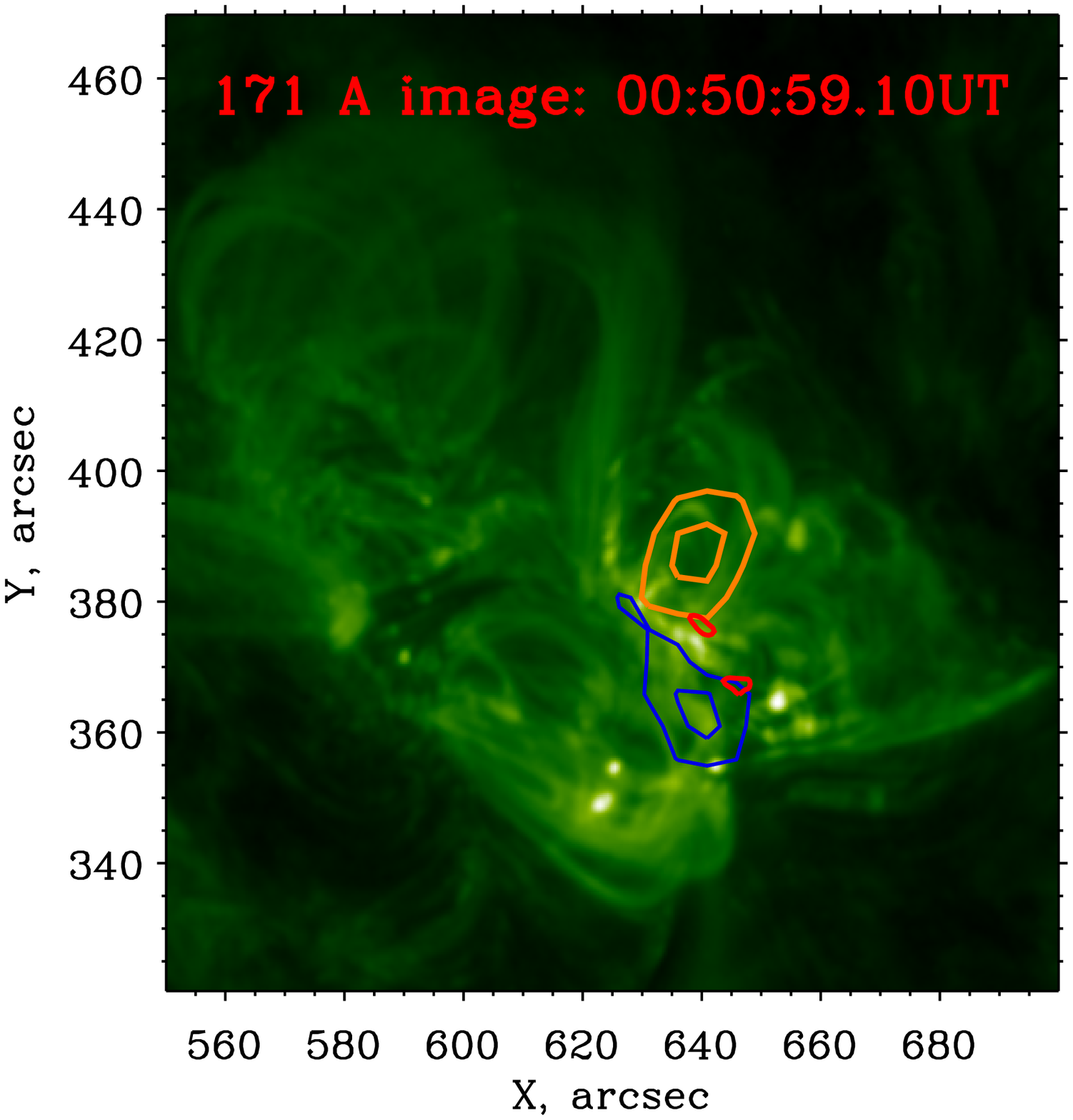}}}
\parbox{0.32\hsize}{
\resizebox{ \hsize}{!}{ \includegraphics{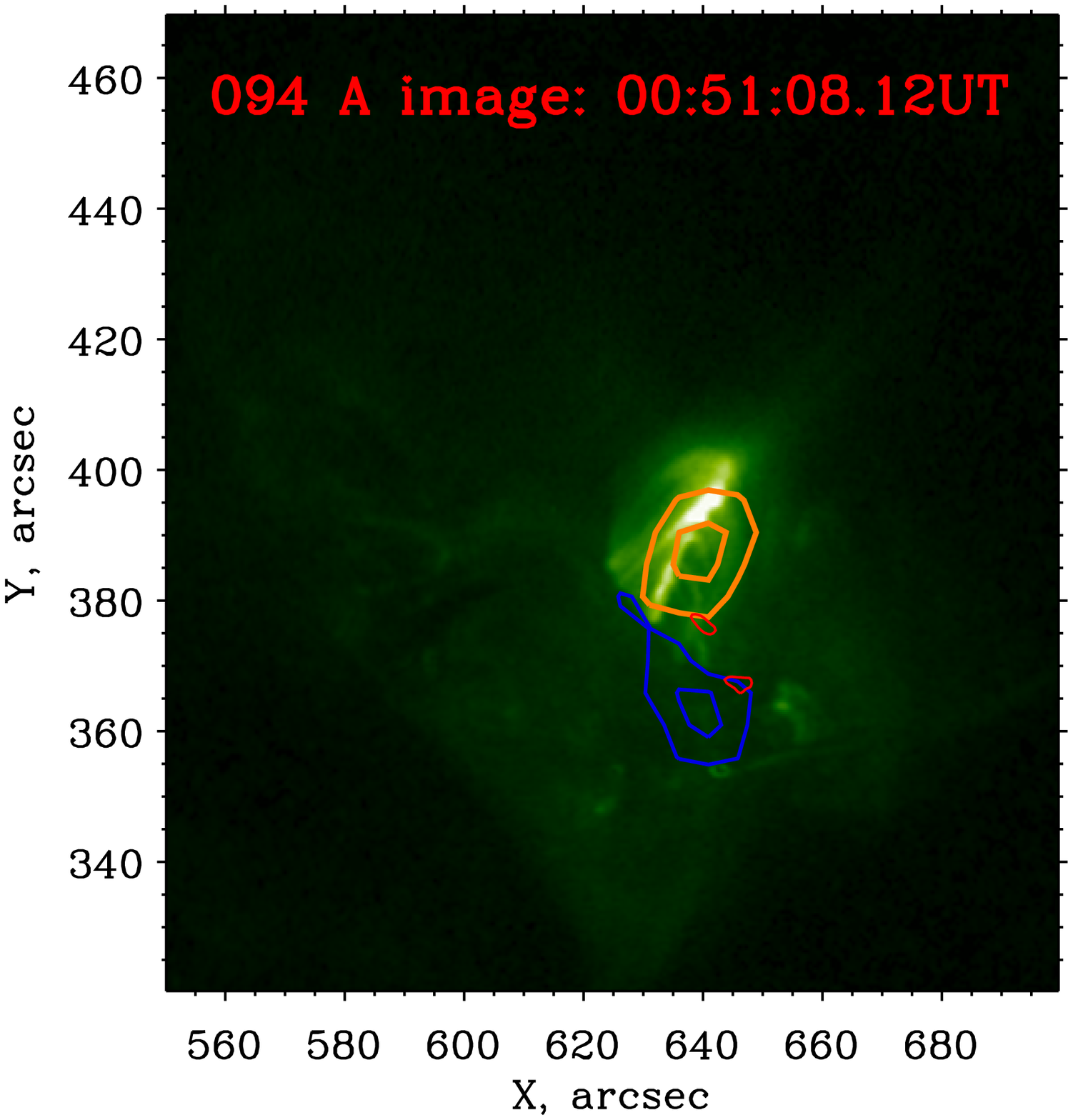}}}
\parbox{0.32\hsize}{
\resizebox{ \hsize}{!}{ \includegraphics{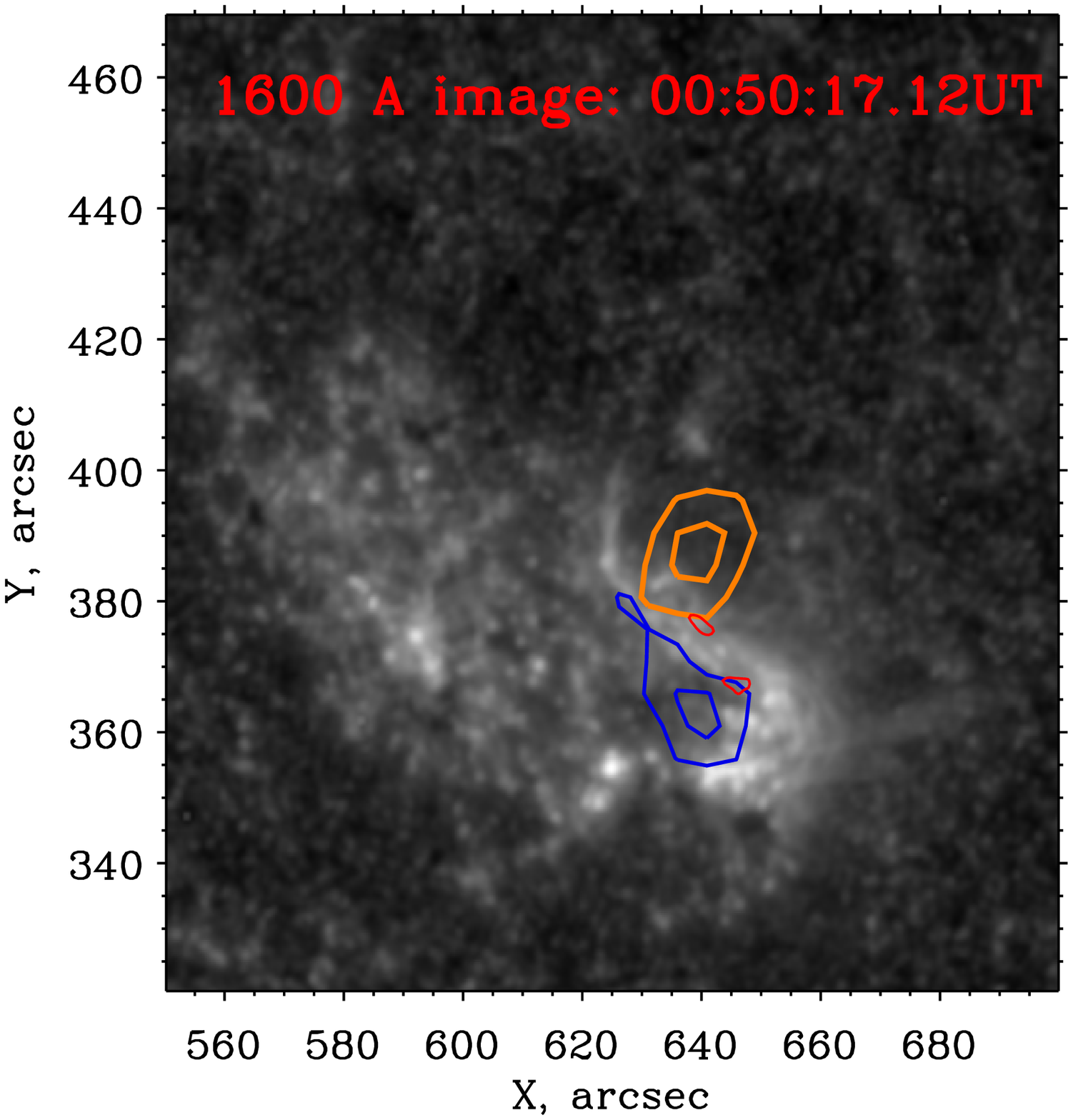}}}

\caption{Position of the MW source before the event onset. The orange contours show the 17 GHz Stokes I at 80$\%$ and 95 $\%$ of the maximum. The blue contours show 17 GHz Stokes V at 50$\%$ and 90 $\%$ of the minumum value. The MW image was taken by NORH at 00:50:03~UT. The red contours show the position of white-light flare kernels (HMI/SDO).}
\label{fig:2}
\end{center}
\end{figure}

As it was noted in Section~\ref{Fl-Observations}, \inlinecite{2011SoPh..tmp..362H} considered the flare precursor onset at about 00:50~UT. The MW intensity and polarization images in 17 GHz reconstructed at 00:51~UT are overlaid on 171~\AA , 94~\AA \ and 1600~\AA \ images at the corresponding times (see Figure~\ref{fig:2}). The 171~\AA \ image did not show any peculiarities.
The MW image taken by NORH at 00:50~UT shows the existence of a quasi-stationary non-polarized MW source at 17 GHz.  As one can see in Figure~\ref{fig:2}(171~\AA~ and 1600~\AA~ images), this source was not associated with the polarized MW spot source. However, the MW intensity and polarization overlaid on the 94~\AA~image indicates that this quasi-stationary MW source is closely connected with the bar in emission seen only in this EUV spectral band.
 Such sources are called Neutral Line associated Sources (NLSs) they are often precursors of powerful flares and are associated with magnetic field separators that are sites of strong energy release in an extended pre-flare current sheet \cite{2008SoPh..249..315U}. One can see in the 94~\AA\ image that the bar in emission is located on the top of the cusp-like structure having footpoints connected to one of the white-light flare kernel. The spatial coincidence of  the NLS and 94~\AA\ bar and their association with the flare kernel indicated that heating and acceleration started at the location where energy could be stored and then propagated to lower layers of solar atmosphere where their manifestation in 1600~\AA\ and white-light were seen.

\begin{figure}
\begin{center}
\parbox{0.32\hsize}{
\resizebox{ \hsize}{!}{ \includegraphics{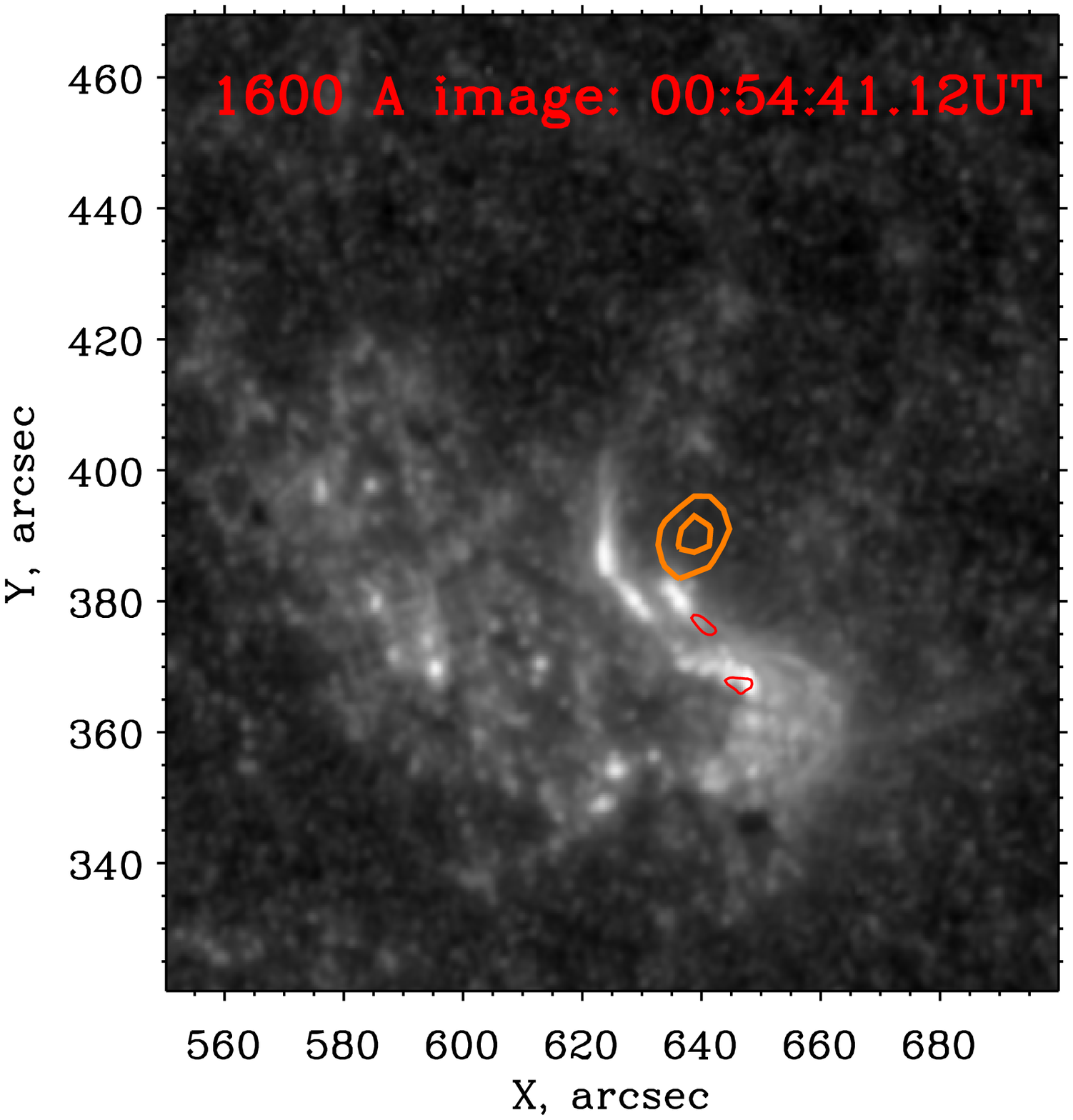}}}
\parbox{0.32\hsize}{
\resizebox{ \hsize}{!}{ \includegraphics{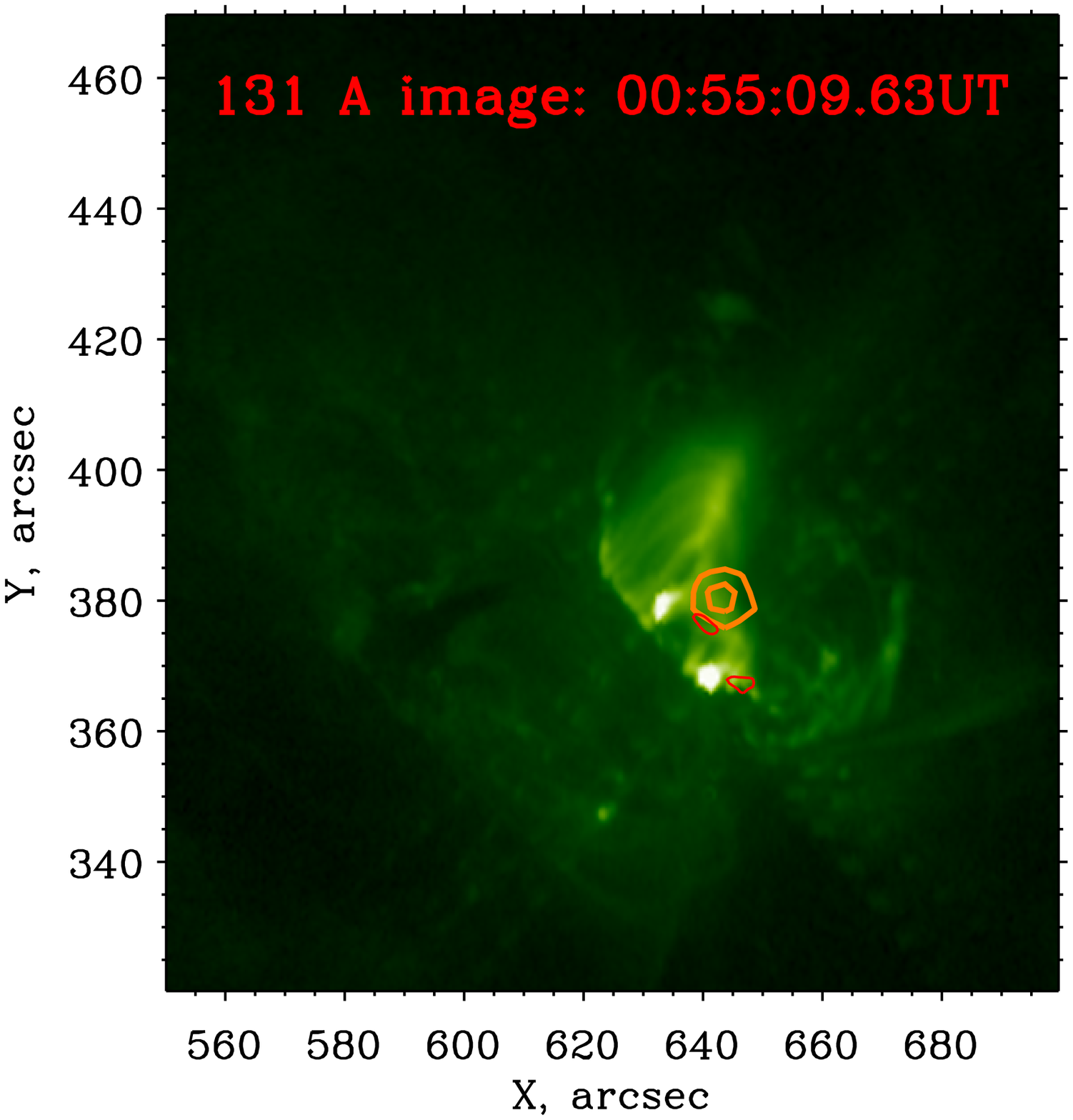}}}
\parbox{0.32\hsize}{
\resizebox{ \hsize}{!}{ \includegraphics{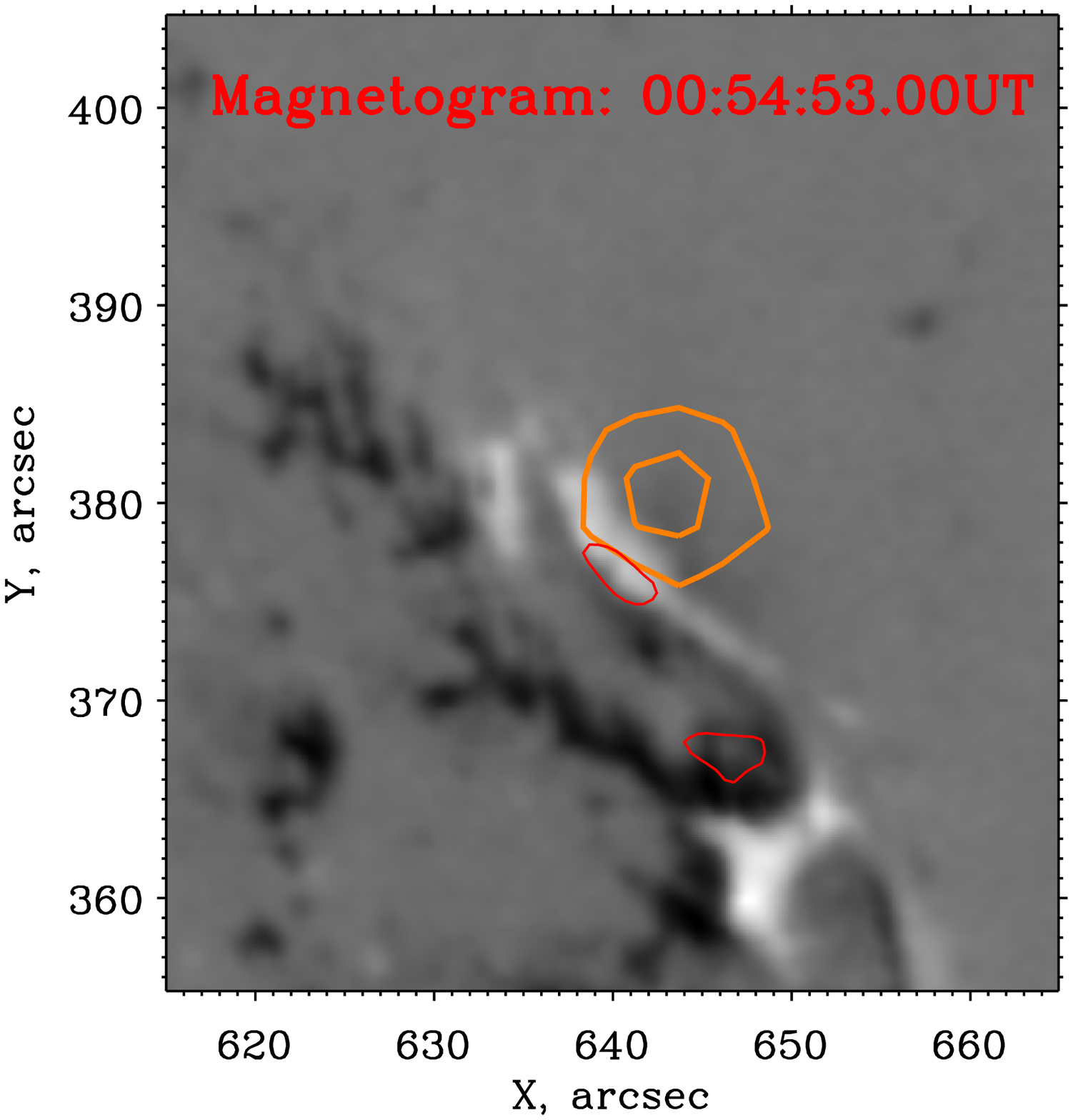}}}

\caption{ Evolution of the MW sources during the precursor. The contours are drawn following the same convention as in Figure~\ref{fig:2}. The MW source image is the closest in time to the EUV (AIA) images and magnetogram (HMI).}
\end{center}
\label{fig:3}
\end{figure}

  At the onset of the flare the 17 GHz MW source moved downward (Figure~\ref{fig:3}).
At this time we see that the flare consists of the two flare ribbons, one flare kernel in the 1600~\AA \ image and two bright kernels in the 131~\AA\ image.
   Only one kernel of the white-light flare was connected both with the flare ribbons and the MW source. Another flare kernel was remote  and located within spots. As it can be seen when comparing to a line-of-sight magnetogram, the flare kernel related to the MW source is located in a region with a very complicated magnetic field structure, while the other kernel lies on a negative polarity feature(about 600 Gauss).

The flare kernel that lies on the negative polarity feature is actually located in the sunspot. According to the results of \inlinecite {2008SoPh..252..149G} and  \inlinecite{2009SoPh..260..135K}, the emission of HXR with flux up to MeV energies, $\gamma$-emission and MW flux at 80 GHz most possibly occurs in the flare kernels located in sunspots. The location of the flare kernel in the sunspot could be one explanation for the presence of HXR emission at high energy  and the MW burst at 80 GHz during the impulsive phase. However, this high energy emission was not simultaneous with the flare onset. Several peaks of non-thermal HXR and MW emission were observed in the studied event before the beginning of the impulsive phase.
So, the peak detected by us in the HXR range of 30-300~keV and in MW emission during initial phase could be the first impulsive peak of this flare. It was weak (about two orders of magnitude lower than the main peak) but it indicated that the event was a flare with a soft-hard-harder (SHH) spectrum evolution \cite{2011SSRv..159..107H}. This fact could explain the contradiction between the apparent impulsive evolution of the event and the difference between MW and HXR spectral indices that would classify this flare as a non-impulsive event according to the results of \inlinecite{2000ApJ...545.1116S}.

\section{Conclusions}
     \label{ S-Conclusions}
We found a quasi-stationary non-pola\-rized MW source in the 17 GHz images during the pre-flare phase. Such quasi-stationary MW source is known as a NLS and it is associated with magnetic field separators that are sites of strong energy release in an extended pre-flare current sheet. Moreover in this MW source coincided with the bar in emission that was detected in the 94~\AA \ EUV image, which was the flare precursor seen in the EUV. The detection of the NLS and the EUV flare precursor during the pre-flare phase and their spatial coincidence indicate the initial storage of a large amount of energy before the flare beginning.

 We found a peak in MW and HXR above 30 keV during the initial phase of the 12 June 2010 flare evolution. The close temporal correlation between the MW emission and the HXR flux indicates the presence of accelerated electrons during the initial phase of this flare. Moreover this fact indicates that both MW and HXR emissions were produced by the common population of accelerated electrons.

    The analysis of the spectral evolution of HXR and MW emission also revealed the presence of accelerated electrons during the initial stages of the flare.The MW spectrum indicates that gyrosynchrotron is the emission mechanism, it also shows a gradual hardening. The difference between the HXR and MW electron spectral indices was about 2 at 00:55:10~UT and exceeded 1 at 00:55:51~UT. According to the results of \inlinecite{2000ApJ...545.1116S} a flare with such difference should belong to the group of non-impulsive flares. This fact could characterize the event as non-impulsive in spite of the short duration and impulsive behavior of the HXR light-curve. Taking into account the non-impulsive character of the energy release process during the flare and the direct evidence of trapping seen in Figure~\ref{fig:0}, a plausible explanation of the harder MW electron index  with respect to the HXR electron index is the "trap plus precipitation" model. Another plausible explanation of this discrepancy is the one suggested by \inlinecite{2009SoPh..260..135K}. These authors noted that sometimes
it can be caused by an underestimation of the microwave turnover frequency resulting from inhomogeneities in the MW source.

 HXR spectra showed a hardening of the photon spectral index during the initial phase from 5.7 to 3.3.
 The thermal parameters of the flare plasma  (the electron temperature and the emission measure)  were increasing as the flare developed. But theirs changes were not so significant as changes of the flare plasma parameters characterizing an acceleration process.
  During the main peak at 00:55:51~UT the non-thermal part of the HXR spectrum continued to become harder. In addition, the fitting of the non-thermal part changed from a single power-law to a double power-law. So, the flare showed a SHH behavior. These facts indicate the presence of several episodes of particle acceleration and confirm that the flare is of the non-impulsive type.

%

%

%

%

%
 \begin{acks} We are grateful to the teams of Nobeyama Radio Observatory, SDO, FERMI and RHESSI who have provided free access to their data. The authors thank an unknown referee for useful suggestions.

This study was supported by the Russian Foundation of Basic Research (12-02-91161, 12-02-00173, 12-02-10006). This research was supported by a Marie Curie International
Research Staff Exchange Scheme Fellowship within the 7th European Community Framework Programme. This work was supported by the Ministry of education and science of
the Russian Federation. \textbf{The final publication is available at springerlink.com}

 \end{acks}

%
%
 \bibliographystyle{spr-mp-sola}
\bibliography{lkk_ref}

%
%
%
%

\end{article}
\end{document}